\documentclass[conference]{IEEEtran}

\IEEEoverridecommandlockouts

\def\BibTeX{{\rm B\kern-.05em{\sc i\kern-.025em b}\kern-.08em
    T\kern-.1667em\lower.7ex\hbox{E}\kern-.125emX}}

 \usepackage{url}
 \usepackage{multirow}

\usepackage{cite}
\usepackage{amsmath,amssymb,amsfonts}
\usepackage{algorithmic}
\usepackage{graphicx}
\usepackage{xcolor}
\usepackage{soul}

\newcommand\xor{\oplus}

\graphicspath{{Figures/}{/home/bman/dev/logos/}}%

\begin{document}

\title{eGPU: A 750 MHz Class Soft GPGPU for FPGA}

\author{\IEEEauthorblockN{Martin Langhammer\IEEEauthorrefmark{1}\IEEEauthorrefmark{2} and
George A.~Constantinides\IEEEauthorrefmark{2}}
\IEEEauthorblockA{\IEEEauthorrefmark{1} Programmable Solutions Group, Intel Corporation\\ \IEEEauthorrefmark{2} Electrical and Electronic Engineering, Imperial College London\\
Email: martin.langhammer@intel.com,
g.constantinides@imperial.ac.uk}}

\maketitle

\begin{abstract}
This paper introduces the eGPU, a SIMT soft processor designed for FPGAs. Soft processors typically achieve modest operating frequencies, a fraction of the headline performance claimed by modern FPGA families, and obtain correspondingly modest performance results. We propose a GPGPU architecture structured specifically to take advantage of both the soft logic and embedded features of the FPGA. 

We also consider the physical location of the embedded memories and DSP Blocks relative to the location and number of soft logic elements in order to have a design with balanced resources. Our goal is to create a high performance soft processor able to implement complex portions of FPGA system designs, such as the linear solvers commonly used in wireless systems, through push-button compilation from software.

The eGPU architecture is a streaming multiprocessor (SM) machine with 512 threads. Each SM contains 16 scalar processors (SP). Both IEEE754 FP32 and INT32 integer arithmetic are supported.  We demonstrate a single SM eGPU in an Intel Agilex device, requiring 5600 ALMs and 24 DSP Blocks, which closes timing at over 770 MHz from a completely unconstrained compile. Multiple eGPUs can also be tightly packed together into a single Agilex FPGA logic region, with minimal speed penalty. 
\end{abstract}


\section{Introduction}

Soft processors have been a longstanding feature of the FPGA landscape~\cite{Microblaze,NiosV}. Typically these have been low performance (both in terms of operating frequency and processing capability), and have been used for handling ancillary functions, or in some cases to control larger datapath structures implemented in the FPGA fabric. Many FPGAs also include embedded processors, typically ARM based~\cite{Zync,AgilexARM}, to support higher performance requirements. Often only a small fraction of the computation is done in the hard processor (almost all of the processing is offloaded to the soft logic array)~\cite{Blott,Suhaib}. In some cases the entire FPGA acts as a processor, for example Microsoft Project Brainwave ~\cite{MicrosoftISCA}.

Integration and flexibility are the key value propositions of the FPGA. While there are many discrete processors that can outperform soft processors, the ability to tightly integrate peripherals and accelerators with a soft processor can often give the FPGA an overall performance advantage. Certain features of the FPGA are already hardened ({\em e.g.}~the DSP Blocks), meaning that FPGAs~\cite{AgilexAGF027} and GPUs~\cite{A100} now support similar levels of floating point density at similar process nodes. This suggests that a well-crafted soft GPU could achieve similar performance density to a hard GPU; by leveraging flexibility in data movement and custom processing elements, the FPGA could outperform it in a system setting. 

The eGPU is designed from the outset to be a high performance soft processor. There is often a large gap between the maximum clock frequency of the FPGA device and the actual speed achieved for a complex design. An FPGA has a natural speed limit, which is governed by the slowest feature, such as the  clock network, embedded memories, or DSP Blocks. In practice, the critical path is usually in the soft logic portion of the design. With our focus on effective design of the soft logic, it is the Agilex DSP Blocks configured in FP32 mode that are the limiting factor for the eGPU at 771 MHz. The combination of a simple memory hierarchy and  shallow processing pipeline of the eGPU provide a low latency and computationally efficient processor for signal processing algorithms that are commonly used in FPGA systems. 

Our eGPU makes the following novel contributions:
\begin{enumerate}
\item Performance: eGPU closes timing at 771 MHz in a current Agilex FPGA without any synthesis, placement, or timing constraints. This is a higher Fmax than any other soft processor we are aware of, of any complexity. 
\item Resource Efficiency: the resource balance of eGPU -- logic, DSP, and memory -- is approximately the same ratio as found in the FPGA. Multiple instances can be specified while maintaining a high efficiency design. In addition, the overall resource requirements are considerably smaller than previously published soft GPUs. 
\item Flexible ISA: eGPU can target a subset of the initialized threads on an instruction by instruction basis. Processing efficiency for operations such as reduction are boosted without requiring thread divergence.
\end{enumerate}

Our goal is not to replace standard GPGPUs, or to compete directly with them, but rather to use the SIMT architecture as a basis for an efficient and effective component of FPGA system design, implementing some algorithms that are challenging to code and maintain in RTL. We are building a different type of GPGPU, one that is works well for small datasets, in terms of both processing efficiency and latency.

\section{Comparison to Other Soft GPU Architectures}
\label{sec:background}

A number of soft GPU architectures have been published, including Guppy~\cite{Guppy}, FGPU~\cite{FGPU}, FlexGrip~\cite{FlexGrip} and MIAOW~\cite{MIAOW}. The capabilities of FGPU and MIAOW have been improved by others in DO-GPU~\cite{DOGPU} and SCRATCH~\cite{SCRATCH}, but at the cost of considerable additional resources and/or Fmax reduction. Of all the previous soft GPU architectures we surveyed, only the FGPU (and its derivatives) were faster than 100 MHz.

Vector processors have also been studied for FPGA, including VEGAS~\cite{Vegas}, VENICE~\cite{Venice}, and VectorBlox~\cite{VectorBlox}, the last of which has been commercialized by Microsemi~\cite{MicrosemiVectorBlox}. The Fmax of these is still modest, with the fastest configuration running at 154 MHz on a recent device. 

FlexGrip and FGPU are compared with our proposed eGPU in Table~\ref{tab:table_softgpus}. This comparison is only representative; there are significant differences between eGPU and the other soft GPUs. FGPU and Flexgrip support enhanced features (like thread divergence), and have a much more complex memory system, with caches. Both can issue instructions across multiple SMs; for eGPU this must be manually controlled by an external agent. But FGPU and FlexGrip have much deeper pipelines than eGPU. While the pipeline depth of eGPU is only 9 (for both Integer and FP operations), FGPU requires 21 cycles for INT and 44 for FP.
However, eGPU is an order of magnitude smaller compared to FlexGrip, and nearly an order of magnitude faster. Although FGPU and FlexGrip are mapped onto older FPGAs (28nm and 40nm planar nodes, respectively), this would not explain the performance difference, especially with the deep pipelines on those processors. When FGPU was ported to a newer Intel Stratix 10 (14nm FinFET process)~\cite{DOGPU}, the clock frequency remained approximately the same.

\section{Architecture}

\begin{figure}[t]
    \begin{center}
        \scalebox{0.6}{\includegraphics{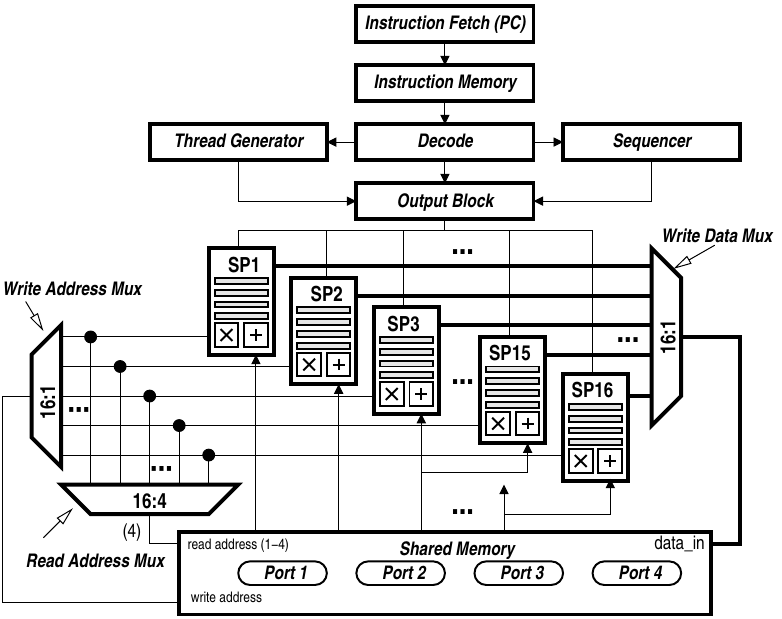}}
        \caption{Representative SM Architecture.}
        \label{fig:SM_Top} 
    \end{center}
\end{figure}

The eGPU is organized in a typical SIMT machine topology, but is designed from the start for an FPGA target. 

We will use the term {\em wavefront} to denote a number of parallel threads issued in a single clock cycle, and {\em thread block} to denote the number of wavefronts required to run all of the initialized threads. Our flexible ISA feature allows both the wavefront and thread block to varied on an instruction by instruction basis.

Each SM (see Figure~\ref{fig:SM_Top}) contains 16 SPs, and additional processing structures, such as dot product operators and special function units (SFUs) can be optionally added. The dot product performs a FP32 vector multiplication and reduction across an entire wavefront, which is supported directly by DSP Block features. The SFU used here is a FP32 inverse square root. Routing fan-out and resource balancing were the main considerations in the choice of the 16 SP per SM. 
The instruction section is not shown, but will be described later in this section. The read bandwidth to the shared memory is four ports, with one write port back from the SPs. Global access is by direct writes into, and reads out of the shared memory at one 32-bit word wide lane.
The eGPU supports a maximum of 512 threads, and has a fixed 16 registers per thread. (We chose these values as they would fit into a single M20K memory, and the were not too dissimilar to older Nvidia GPGPUs~\cite{NvidiaFermi}). 
Currently, a 2D thread space can be defined. All data, whether floating point or integer, is 32 bits wide. Correspondingly, wavefronts are 16 wide, and a thread block is up to 32 wavefronts deep. Load (memory and immediate), store, and processing (both FP32 and INT32) have different latencies. Hazards have to be managed by the programmer; there are no hardware interlocks. These hazards, however,  (especially dependencies inside the SP) are typically only exposed for small thread blocks.

\subsection {Shared Memory}

The shared memory is configured as a four read port, one write port memory, which requires 4 identical copies of the shared memory space to be maintained. The four read ports are transferred to the 16 SPs in a four phase sequence. Writeback is a 16 phase sequence. The shared memory bandwidth, especially the single word store, is one of the most significant performance bottlenecks for the eGPU. Later in this section, we will describe some novel architectural features which mitigate these limitations. 

\begin{table}
\small
  \begin{center}
    \caption{Resource Comparison.}
    \label{tab:table_softgpus}
    \begin{tabular}{c|c|c|c|c} 
       \textbf{Architecture} & \textbf{Configuration} & \textbf{LUTs}& \textbf{DSP} & \textbf{FMax}  \\
      \hline
      \textbf{FGPU~\cite{FGPU}} & {2CUx8PE} & {57K} & 48 & 250 \\
        \hline
      \textbf{FlexGrip~\cite{FlexGrip}} & {1SMx16PE} & {100K} & 300 & 100  \\
        \hline
      \textbf{eGPU} & {1SMx16SP} & {5K} & 24 & 771  \\
      \hline
    \end{tabular}
  \end{center}
\end{table}

\subsection {SP Architecture}

\begin{figure}
    \centering
    \includegraphics[scale=0.65]{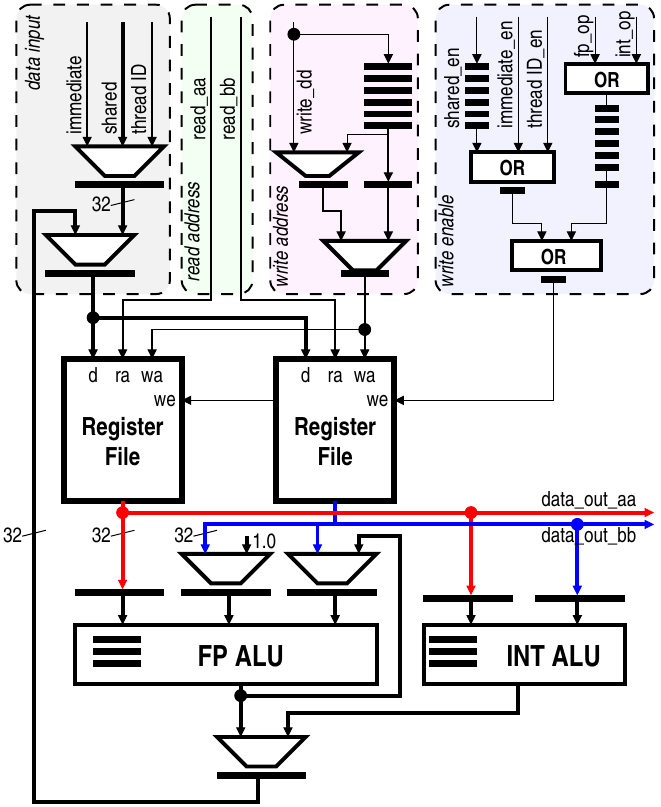}
    \caption{SP Architecture.}
    \label{fig:SP1}
\end{figure}

The architecture of the SP is shown in Figure~\ref{fig:SP1}. A register file with a total of 512 32-bit words is configured as a two read port, one write port memory. The SP can read from, and write to, shared memory, and execute both IEEE754 single precision floating point (FP32) and 32-bit integer (INT32) instructions. We expect that most of the work will be done by the FP ALU, and the INT ALU will mostly be used for address generation. At this time predicates (a conditional branch on a thread by thread basis) are not supported, as we have found that the benchmarks we are most interested in (such as FFT and matrix decomposition) do not make data dependent decisions. All SPs have the same micro-architecture, except the first lane, as the dot product and SFU cores write into this lane. The FP ALU uses the DSP blocks, and does not require any soft logic other than an input multiplexer to convert the pre-configured FP32 multiply-add datapath into an adder. 

The INT ALU is constructed out of ALMs, plus half a DSP Block for the INT multiply. The multiply is 16x16 with a 32-bit output, which will typically be used for address generation. In addition to the multiplier, the ALU contains logic functions (AND/OR/XOR/NOT), add/sub, and shifters, all with a  pipe depth to match the FP ALU. Some integer functions (especially the shifters and the add/sub) could restrict the 770 MHz goal when in a full chip or otherwise densely packed design, but the added logic depth here allows us to spread some of these INT32 functions over two pipeline stages. As an example, the adders are implemented using a carry select structure.

\subsection {Instruction Section}

The instruction section consists of an instruction fetch, instruction memory, instruction decoder, sequencer and thread generator.

The instruction fetch stage also supports zero-overhead loops, similar in style to DSP processors~\cite{TI55X}. A loop counter is initialized by instruction (which requires only a single cycle) and is automatically decremented at the bottom of each loop (another single cycle instruction). 

\begin{figure}
    \centering
    \includegraphics[scale=0.40]{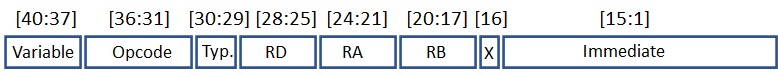}
    \caption{I-WORD.}
    \label{fig:iword}
\end{figure}

The instruction memory (I-MEM) contains 40-bit wide words, and has a parameterizable depth. The program size for the expected applications is relatively small. A single M20K can hold a 512x40 block, so we expect that one to four memories would likely be sufficient, although there is no upper limitation on the I-MEM size. (Two of the benchmarks in this paper, the 256 point radix-2 FFT and the 16x16 QRD require 135 and 40 instructions, respectively, which suggest that multiple programs can easily be contained in a single M20K). The I-MEM can be independently reloaded or updated, including during execution of a program - the external agent just has to be aware of the memory space and update a portion of the I-MEM not currently being accessed.

The sequencer controls the application of the decoded instruction signals to the SPs. Some instructions are single cycle, but most require multiple cycles. Operation instructions (FP or INT) typically run as many wavefronts until all threads have been executed. Load and store instructions similarly run for all threads, but loads require one clock per four threads loaded, and stores take one cycle per thread. The sequencer also adjusts the number of wavefronts with the variable thread block capability explained below.

\begin{table}
\footnotesize
  \begin{center}
    \caption{Instruction Set.}
    \label{tab:table_instructions}
    \begin{tabular}{|c|c|c|}
    \hline
  \textbf{Group} & \textbf{Instruction} & \textbf{Comments} \\
      \hline
      \multirow{3}{*}{Arithmetic} & {ADD.TYPE Rd,Ra,Rb} & {Rd = Ra + Rb} \\
    \ & {SUB.TYPE Rd,Ra,Rb}  & {Rd = Ra - Rb} \\
    \ & {MUL.TYPE Rd,Ra,Rb}  & {Rd = Ra * Rb} \\
    \hline
    \multirow{4}{*}{Logic} & {AND Rd,Ra,Rb}  & {Rd = Ra \& Rb} \\
    \ & {OR Rd,Ra,Rb}  & {Rd = Ra $\|$ Rb} \\
    \ & {XOR Rd,Ra,Rb}  & {Rd = Ra $\xor$ Rb} \\
   \ & {NOT Rd,Ra,Rb} & {Rd = \!Ra}\\
    \ & {LSL Rd,Ra,Rb} & {Rd = Ra $\ll$ Rb} \\
    \ & {LSR Rd,Ra,Rb} & {Rd = Ra $\gg$ Rb} \\
    \hline
    \multirow{2}{*}{Memory} & {LOD Rd (Ra)+offset} & {Read from Shared}\\
    \ & {STO Rd (Ra)+offset} & {Write to Shared} \\
    \hline
    \ {Immediate} & {LOD Rd \#Imm} & {Rd = Imm} \\
    \hline
    \multirow{3}{*}{Thread} & {TDx Rd } & {Rd = Thread IDx}\\
    \ & {TDy Rd} & {Rd = Thread IDy}\\
    \hline
    \multirow{3}{*}{Extension} & {DOT Rd,Ra,Rb} & {Dot Product $\langle Ra,Rb \rangle$} \\
    \ & {SUM Rd,Ra,Rb} & {Reduction $\langle Ra,Rb \rangle$}\\
    \ & {INVSQR Rd,Ra} & {$Rd = 1/\sqrt {Ra}$} \\
    \hline
    \multirow{3}{*}{Control} & {JMP address} & {Jump to Address}\\
    \ & {JSR address} & {Subroutine Address}\\
    \ & {RTS} & {Return from Subroutine}\\
    \ & {LOOP address} & {Jump and Dec Loop Ctr} \\
    \ & {INIT loops} & {Set Loop Ctr}\\
    \ & {STOP} & {Stop and Set Flag}\\
    \hline
    \end{tabular}
  \end{center}
\end{table}

The 40-bit instruction word is divided into eight fields. The most significant field is the 4-bit {\tt Variable}, which can modify the thread block in real time. This field is described later in this section. Next is the 6-bit {\tt Opcode} field. This allows for 64 instructions, and we currently implement 23 (see Table~\ref{tab:table_instructions}). The 2-bit {\tt Type} field selects INT32, UINT32, or FP32 numerics when an operation is run. The next three fields, 4-bits each, select the destination and two source registers, or alternately the address register in case of an indexed load or store. The single bit {\tt X} field enables thread snooping, explained below. The 15-bit immediate field is sign extended to 32 bits. It also contains the register address extensions when thread snooping is selected.

\subsection {Novel Architectural Features}

The flexible ISA enables the thread block -- in both depth and width -- to be changed on an instruction by instruction basis, while keeping a constant thread initialization. No data flush is required -- the switch occurs instantly with no latency, other than to manually control hazards as the depth changes. This effectively allows the eGPU to switch personalities between SIMT, Vector Processor (VP), multi-threaded CPU, and simple MCU architectures at any point. This can significantly reduce the negative impact of low memory bandwidth. We will see the effect of this in some of the benchmarks, especially in matrix decomposition, where a normalization value is calculated by a dot product. The result of the dot product will be written back to only the first lane. Through the ability to only save a single thread per wavefront, or even a single thread within a block, a large number of cycles is saved. This feature is controlled by the upper 4 bits of the I-Word. Bits [40:39] set the width of the wavefront (full width, 1/2 width, 1/4 width, and single thread), and bits [38:37] set the depth of the block (full depth, 1/2 depth, 1/4 depth, and single cycle). 

The Modified Gram-Schmit (MGS) QRD algorithm demonstrates the power of this variable ISA. Our benchmark is on a 16x16 matrix, which uses 256 threads. A norm is computed on a single column, and then applied to all remaining columns. First, a single wavefront is isolated, to compute the norm. Then a single thread is isolated to write the norm into shared memory, from where it can be broadcast to all threads. A regular SIMT instruction is issued to apply the norm to the remaining columns. With a standard GPU architecture, thread divergence would be used to isolate subsets of the initialized thread space, requiring the running of all threads (whether or not they were executed). One of the limitations of the eGPU is the writeback bandwidth; but using the flexible ISA, the norm writeback only requires a single clock cycle.

The VP mode also uses the thread snooping feature. When the {\tt X} bit in the instruction word is set, two 5-bit sub-fields in the immediate section of the I-Word provide the upper five bits of the operation source registers. This allows the threads of the first wavefront (threads 0 to 15) to access any register in their lane. An example where thread snooping can be used is in the reduction benchmark. The dot product operator writes the reduction of each wavefront into the first lane (threads within the first SP). The first thread of the first lane can then access all of these threads directly, without having to go through the shared memory.

\subsection {FPGA Mapping}

One of our design goals was that the resource usage of the eGPU was balanced. The most common sector type on Agilex devices contains 237 M20Ks, 164 DSP Blocks, and 16,400 ALMs \cite{ChromczakAgilex}. All of these resources are arranged in columns, 41 LABs high; the mix of resources can be seen in the floorplan output by Quartus. As most of the soft logic is in the INT ALU, we can trade off various feature sets in this block to enable fitting into specific geometries, such as sector boundaries. We will now consider some possibilities for a balanced design inside a sector. Our base eGPU architecture needs 24 DSP Blocks (16 for the FP ALU and 8 for the INT ALU). Each SP uses two M20Ks for register files (32 in total). To pack four SMs per sector requires 128 M20Ks and 96 DSP blocks. This leaves 109 M20Ks remaining for shared memory, which is 27 512x32 memories per eGPU. For a quad read port shared memory, this allows us a 6 deep (3K word) shared memory, or 12K bytes. There are 68 DSP Blocks remaining, or 16 per eGPU, which is how many DSP Blocks are required to implement the dot product core. The sector contains 1640 LABs (16400 ALMs), which gives us a budget of 4100 ALMs per eGPU. 

\section{Benchmarks}

We demonstrate the performance and utility of the eGPU by running two non-trivial benchmarks, FFT and QRD. We also profile the code to provide an analysis of the strengths and weaknesses of the eGPU architecure, and to provide a starting point for future architectural enhancements.

\subsection {FFT}

\begin{table}
\small
  \begin{center}
    \caption{FFT Profile (256 Points).}
    \label{tab:fft_profile}
    \begin{tabular}{c|c|c|c} 
       \textbf{Instruction Type} & \textbf{Cycles} & \textbf{\%}& \textbf{Comment} \\
      \hline
      \textbf{LOD Immediate} & 64 & 5 & {Set up Address Generation} \\
        \hline
      \textbf{Logic} & 48 & 4 & {Address Generation} \\
        \hline
      \textbf{INT} & 32 & 3 & {Address Generation} \\
      \hline
      \textbf{LOD Indexed} & 384 & 32 & {Get Data, Twiddles} \\
      \hline
      \textbf{FP32 Add/Sub} & 96 & 8 & {Butterfly} \\
      \hline
      \textbf{FP32 Multiply} & 64 & 5 & {Butterfly} \\
    \hline
      \textbf{STO Indexed} & 512 & 43 & {Writeback Pass to Shared} \\
      \hline
    \end{tabular}
  \end{center}
\end{table}

We coded a radix-2 (R2) decimation-on-frequency (DIF) FFT, and analyzed the performance for lengths 32 and 256. The R2 butterfly takes two complex inputs. They are added for the upper output. The lower output is the subtraction of the two inputs, followed by a rotation, implemented as a complex multiply by a coefficient. Here, we map each butterfly to its own thread. The 32 point FFT therefore requires only 16 threads, which maps to a single wavefront of the eGPU. The 256-point FFT requires eight wavefronts, which is slightly less than the pipeline depth of the SP. This creates a RAW hazard at one point in the address generation code, which we handle by inserting a NOP. We will see that the shared memory bandwidth limitation is the most significant performance limitation for this benchmark. 

As an example of the assembly code style, the following code segment calculates the address for each thread at the start of each pass ({\tt R1} contains the threadID, {\tt R3} and {\tt R4} contain address masks, {\tt R5} the address rotate value - always `1' for radix-2, and {\tt R9} the twiddle addressing increment for this pass). As the eGPU is a SIMT machine, this instruction runs for all active threads, with the threadID creating a unique address for each one.  We will follow the execution for thread ID 110 ({\tt "01101110"}) for the 256 point FFT in pass 2 as an example. The initial values are {\tt R3 = "01000000"}, \linebreak {\tt R4 = "00111111"}, {\tt R5 = 1}, and {\tt R9 = 2}. The resultant data address is 174 ({\tt "10111000"}).

\begin{verbatim}
AND.INT32 R6,R1,R3;	// R6 = "01101110”
AND.INT32 R7,R1,R4;	// R7 = "00101110”
LSL.INT32 R8,R6,R5;	// R8 = "10000000”
ADD.INT32 R6,R7,R8; // R6 = "10101110"
NOP;                // prevent RAW hazard
ADD.INT32 R2,R6,R6; // R2 = "101011100"
LSL.INT32 R3,R7,R9; // R3 = "0010111000”
RTS
\end{verbatim}

Table~\ref{tab:fft_profile} shows the distribution of the instruction types for a radix-2 FFT. The address calculation takes 12\% of the cycles, and the actual butterflies 13\%, with the shared memory access accounting for 75\% of the total. Every pass needs to go through the shared memory so that the butterflies obtain the correct input data locations. The combination of the low memory bandwidth for the eGPU and the number of passes for the R2 FFT makes this example relatively inefficient. 

\subsection {QRD}
\begin{table}
\small
  \begin{center}
    \caption{QRD Profile}
    \label{tab:qrd_profile}
    \begin{tabular}{c|c|c|c} 
       \textbf{Instruction Type} & \textbf{Cycles} & \textbf{\%}& \textbf{Comment} \\
      \hline
      \textbf{NOP} & 44 & 15 & {Hazards} \\
        \hline
      \textbf{INT} & 16 & 5 & {Address Generation} \\
        \hline
      \textbf{LOD Indexed} & 132 & 44 & {Broadcast Values} \\
      \hline
      \textbf{FP32 Add/Sub} & 16 & 5 & {Apply Projection} \\
      \hline
      \textbf{FP32 Multiply} & 32 & 10 & {Apply Projection} \\
            \hline
      \textbf{FP32 Dot} & 17 & 6 & {Calculate Norm} \\
            \hline
      \textbf{FP32 SFU} & 1 & 1 & {Calculate Norm} \\
    \hline
      \textbf{STO Indexed} & 33 & 11 & {Write Q,R to Shared} \\
      \hline
    \end{tabular}
  \end{center}
\end{table}

We also coded a QRD using the Modified Gram-Schmidt algorithm~\cite{demmel97}, and show results for a small 16x16 matrix. This size of matrix would perform particularly poorly on standard GPUs~\cite{KerrQR}. The combination of the dot product core and the SFU make the calculation of the norm value (in the outer loop) very quick. The ability to specify the execution of both a subset of the thread space and/or the subset of a wavefront means that indexed store operations may need to run for as little as a single clock cycle. 

The results can be seen in Table~\ref{tab:qrd_profile}. The FP operations (actual QRD calculation work) are 22\% of the total cycles, and writeback to the shared memory is only 11\%. Interestingly, the broadcast of the norm from a single thread to all threads (which needs to go through the shared memory) requires almost half of the total time, which is a potential focus area for future architectural optimizations. The hazard mitigating NOPs require 15\% of the cycles, but these would largely disappear for larger matrices. The QR decomposition of these smaller matrices on hard GPUs, even with vendor libraries such as cuBLAS, often have efficiencies measured in the small single digits~\cite{Dongarra}. 
But when the actual number of arithmetic operations per instruction is taken into account, the true efficiency of the eGPU is much higher. The FP32 dot product runs for 6\% of the cycles, but each dot product instruction performs 31 operations (16 multiplies and 15 adds). This structure is particularly effective on the type of operations needed for MGS, while most of the hard GPU examples we surveyed~\cite{KerrQR,Dongarra} used Householder reflections~\cite{GolubVanLoan96}.

\section{Results}

\begin{table}
\small
  \begin{center}
    \caption{Resource Report}
    \label{tab:table_resources}
    \begin{tabular}{c|c|c|c|c} 
       \textbf{Block} & \textbf{ALM} & \textbf{Registers}& \textbf{DSP} & \textbf{M20K} \\
      \hline
      \textbf{Instruction} & 235 & 540 & 0 & 2 \\
        \hline
      \textbf{SM} & 5372 & 14996 & 24 & 48 \\
        \hline
      \textbf{SP} & 267 & 794 & 1.5 & 2 \\
      \hline
      \textbf{INT ALU} & 114 & 249 & 0.5 & 0 \\
      \hline
    \end{tabular}
  \end{center}
\end{table}

We performed a completely unconstrained compile of the eGPU into an Intel Agilex AGFB014R24A1E1V device using Quartus 22.4.0 Pro Edition.
The resource requirements are shown in Table~\ref{tab:table_resources}.  This closes timing at 771 MHz, with the DSP Block in FP32 Multiply-Add mode the critical path. The other failing paths are primarily soft logic inside the integer ALU. Additional experiments with additional pipelining and alternate constructions of the INT ALU achieve only marginal gains, with a soft logic Fmax of 831 MHz. As a result, we can be confident that our core is performant by design {\em i.e.}~it will compile to the maximum performance (limited by the DSP Block) without any constraints. 

As a result of the resource ratio-driven design methodology, we are able to replicate the eGPU core four times, in a spatially regular fashion, by locking the design into a sector boundary. The logic is slightly larger than the contents of the sector, so about 20\% of this design spilled over into the next sector in the Y direction, although still within the horizontal sector boundaries. The same process allows placement of any number of tightly packed quad-eGPU structures side by side if needed, without any reduction in performance on a single quad construct. (There is no routing spillage over the boundaries horizontally). There is a slight ($\sim5\%$) performance degradation in the quad packing to 738 MHz.

\section{Conclusions}
This paper describes the architecture and implementation of a high performance embedded GPU, which can achieve over 770 MHz operating frequency with an unconstrained compile. The area is modest, and has a  balanced resource usage of memories, DSP Blocks, and soft logic. As a consequence, we can also tightly pack multiple cores together with a minimal performance degradation. The flexibility of the FPGA means that we can optionally add accelerator cores such as dot-product cores and operators such as elementary function cores. We have also coded and demonstrated benchmarks such as FFTs and QR matrix decomposition. 

\newpage\clearpage


\end{document}